\documentclass[12pt]{article}
\usepackage{graphicx}
\usepackage{amsmath}
\usepackage{longtable}

\begin{document}

\makeatletter
\renewcommand*{\@cite}[2]{{#2}}
\renewcommand*{\@biblabel}[1]{#1.\hfill}
\makeatother

\title{Red Giant Clump in the Tycho-2 Catalogue}
\author{G.~A.~Gontcharov\thanks{E-mail: georgegontcharov@yahoo.com}}

\maketitle

Pulkovo Astronomical Observatory, Russian Academy of Sciences, Pul\-kov\-skoe sh. 65, St. Petersburg, 196140 Russia


The Tycho-2 proper motions and Tycho-2 and 2MASS photometry are used to select 97348
red giant clump (RGC) stars. The interstellar extinction and photometric distance are calculated for each
of the stars. The selected stars are shown to form a selection-unbiased sample of RGC stars within about
350 pc of the Sun with the addition of more distant stars. The distribution of the selected stars in space and
their motion are consistent with the assumption that the RGC contains Galactic disk stars with various
ages and metallicities, including a significant fraction of stars younger than 1 Gyr with masses of more
than 2 $M_{\odot}$. These young stars show differences of their statistical characteristics from those of older RGC
stars, including differences in the variations of their distribution density with distance from the Galactic
plane and in the dispersion of their velocities found using radial velocities and proper motions. The Sun
has been found to rise above the Galactic plane by $13\pm1$ pc. The distribution density of the stars under
consideration in space is probably determined by the Local Spiral Arm and the distribution of absorbing
matter in the plane of the Gould Belt.

\newpage
\section*{INTRODUCTION}

A region of enhanced distribution density of red giants called the red giant clump (RGC) is clearly
seen on the color--absolute magnitude diagrams for open clusters and various samples of stars. Since
these stars have a specific absolute magnitude, their photometric distances can be determined rather accurately.

In globular clusters, the horizontal branch (HB) occupies approximately the same region of the color--absolute
magnitude diagram. However, in contrast to the HB, the RGC is not an evolutionally homogeneous
group of stars and consists of three groups considered below that are located in this region of the
diagram for various reasons.

(1) It is traditionally believed that low-mass core helium burning stars after the red giant branch
(RGB) phase and a helium flash constitute an overwhelming majority of the RGC (and HB) stars
(Girardi et al. 1998). Before the flash, the mass of such a star is lower than some limiting value,
M$_{HeF}\approx$2 M$\odot$, and the helium core is \emph{degenerate}. After the flash, the degeneracy is removed and the
luminosity of these helium burning stars is almost independent of the color index and age, since the
helium core mass at this time is almost independent of the stellar mass. However, the color index depends
on the mass, age, and metallicity: it decreases with decreasing metallicity and decreasing age (increasing
mass) (Girardi 1999). Nevertheless, if the metallicity has increased appreciably over the lifetime of the
Galactic disk, these dependences almost cancel each other out and the stars under consideration should
have a moderately large spread in color (Girardi et al. 1998). The main-sequence (MS) progenitors of these
stars are F and early-G stars, because later-type stars could not fall into the RGC over the lifetime of the
Galactic disk ($<10^{10}$ yr). Since these stars are fairly old by the helium flash, below we call them RGC old
stars (RGC-O). A typical Galactic disk star with a mass of 1.2 $M_{\odot}$ and a solar metallicity
($\mathbf Z=0.02$, $\mathbf Y=0.27$) spends $\approx6\times10^{9}$ yr on the MS, $0.4\times10^9$ yr
as a subgiant, $0.6\times10^9$ yr on the RGB, and only $0.1\times10^9$ yr in the RGC (Castelliani et al. 1992;
Girardi et al. 1998). Thus, the clump of these stars in the RGC is formed only because the RGB stars
with a large spread in absolute magnitude and color ($-2<M_V<4$, $0.7^m<(B-V)<1.6^m$) concentrate for
a certain time in the RGC with a small spread ($0.5<M_{V}<1$, $0.8^m<(B-V)<1.2^m$). By comparing these
ranges and the residence times on the RGB and in the RGC, it may be concluded that the distribution
density of the RGC stars should be higher than that of the RGB ones only a factor of 2--3. Such are
the diagrams for globular clusters. In contrast, in the solar neighborhood, i.e., in the Galactic disk,
the RGC is much more densely populated than the RGB. Once the Hipparcos Catalogue (ESA 1997)
has been published, it has become clear that the distribution density of the RGC stars with respect to
the RGB is appreciably higher than that predicted by the theory (Girardi et al. 1998). This can be partly
explained theoretically by the fact that low-mass metal-poor stars fall into the RGC relatively quickly:
a star with M = 1.5 $M_{\odot}$ and $\mathbf Z=0.001$ falls into the RGC $\approx2$ Gyr after its birth
(Girardi et al. 2000). In addition, Girardi et al. (2005) showed that young,
relatively massive stars theoretically could constitute an appreciable fraction of the RGC in spiral arms and
other regions where the stars have been born for the last 1 Gyr.

(2) Thus, high-mass ($M>M_{HeF}$) core helium burning stars after a fairly short hydrogen shell burning
phase of a \emph{nondegenerate} helium core also fall into the RGC. Since these stars are rather young
by helium ignition, below we call them RGC young stars (RGC-Y). Their luminosity depends on the helium
core mass, while the latter, in turn, depends on the stellar mass. However, the fraction of stars with
masses higher than 3 $M_{\odot}$, their total lifetime, and the residence time in the RGC are so small that, in fact,
they are lost in samples with less massive stars and are absent in the RGC (Girardi 1999). As a result,
the spread in luminosity and absolute magnitude for RGC-Y dominated by stars with $M\approx2 M_{\odot}$ is only
slightly larger than that for RGC-O. In addition, being young, RGC-Y have a small spread in metallicity
and, therefore, occupy a compact region of the color--absolute magnitude diagram at the blue edge of the
RGC in all samples of Galactic disk stars (Girardi et al. 1998). It should be noted that RGC-Y concentrate
in the RGC region for a different reason than RGC-O: the core helium burning time for RGC-Y is
considerably longer (about 200 Myr) and is shorter than the residence time of such a star on the MS,
where its type is B or A, only by a factor of several. The ratio of the helium burning time to the hydrogen
burning time and, accordingly, the ratio of the distribution densities of the RGC and MS stars is
at a maximum for stars with a mass of about 2 $M_{\odot}$ (Bressan et al. 1993).

(3) Some of the RGB stars can have the same color and absolute magnitude as RGC-O and RGCY,
although they are at a different evolutionary phase and should be considered as a foreign admixture in the
samples of RGC stars.

The fraction of these three groups in the RGC affects the mean absolute magnitude of the RGC for
various samples of stars and, accordingly, the photometric distances (Girardi et al. 1998). In addition,
many of the observed differences between globular and open clusters are determined by the large fraction
of RGC-Y among the giants of open clusters (Girardi and Bertelli 1998). In clusters younger than 1 Gyr,
the RGC should include only RGC-Y. Analysis of the RGC composition and properties of these stars
is also important in testing the theory of stellar evolution and, in particular, in analyzing the metallicity
dependence of the critical mass $M_{HeF}$ and modeling the helium flash and its consequences. This paper is
one of the stages in investigating the RGC.

The fraction of RGC-Y and RGC-O for the nearest part of the Galactic disk can be estimated
by selecting RGC stars from the Hipparcos and Tycho-2 Catalogues (H\o g et al. 2000) using $J$, $H$, and $Ks$
infrared photometry from the 2MASS Catalogue (Skrutskie et al. 2006) and by analyzing
their distribution in space and velocity dispersion. A similar study of O-B stars using proper motions
and multiband photometry was performed previously (Gontcharov 2008). However, the much smaller
spread in absolute magnitude for the RGC stars allows us to use their proper motions only to select stars
by calculating their distances only from photometric data.

It should be noted that, being magnitude-limited, the Hipparcos and Tycho-2 Catalogues cannot contain
any RGC stars outside the Galactic disk (in the bulge, halo, etc.): the RGC stars brighter than $V=11^{m}$
at an extinction as high as $1^m$ are closer than 1.5 kpc.

\section*{RGC STARS IN THE HIPPARCOS CATALOGUE}

Following Girardi et al. (2005), let us compare the observed and theoretical color--absolute magnitude
diagrams for 16 726 stars from the Hipparcos Catalogue with the most accurate data. The proper motion
components for these stars are known with an accuracy better than 0.002 arcsec yr$^{-1}$, the relative
error in the parallax is less than 10\%, the Hipparcos photometric magnitudes $B_T$ and $V_T$ are known with
an accuracy better than $0.1^{m}$, the stars are neither double nor multiple, and their spectral classification
is known. Since they are no more than 250 pc away, their reddening is small and the absolute magnitudes
calculated from their parallaxes will have no systematic errors. Figure 1a shows the positions of
giants and MS stars of late spectral types among these stars on the $(B-V)$ -- $M_V$ diagram based on
Hipparcos data. Theoretically, according to Girardi et al. (2005), RGC-O with an age from $2\times10^9$
to $10\times10^9$ yr and metallicity $0.004<\mathbf Z<0.020$ should occupy the region marked by the oval (the
more metal-poor and, accordingly, older stars are located leftward); the RGB with the same age and
metallicity should occupy the region between the two dotted lines (again, the more metal-poor and,
accordingly older stars are located leftward); and RGC-Y with an age up to $2\times10^9$ yr and metallicity
$0.006<\mathbf Z<0.020$ should occupy the region marked by the nearly rectangular figure (the more metal-poor
stars are located leftward; no age gradient is seen). The distribution of stars within the shown regions
and the sharp decrease in distribution density outside them confirm the conclusion by Girardi et al. (2005)
that RGC-Y and RGB stars constitute a significant fraction of the RGC stars in the nearest part of the
Galactic disk and are located on the diagram precisely as predicted by the theory. Since the RGC in the
Tycho-2 Catalogue also contains mostly disk stars within the nearest kiloparsec, it can be assumed that
the color -- absolute magnitude diagram for the Tycho-2 data and the conclusions drawn from it are similar
to the results obtained from Hipparcos data.

Having assumed a uniform distribution of the RGC stars in age from $1\pm0.1$ to $10\pm2$ Gyr and
in metallicity $\mathbf Z$ from $0.004\pm0.001$ to $0.020\pm0.001$
(Bertelli et al. 2008; Girardi and Salaris 2001; Salaris and Girardi 2002), we calculate the theoretical mean
absolute magnitudes of the RGC stars and their uncertainties due to these variations in initial parameters:
$\overline{M_{B_{T}}}=2.06^{m}\pm0.05^{m}$,
$\overline{M_{V_{T}}}=0.86^{m}\pm0.05^{m}$,
$\overline{M_{J}}=-0.87^{m}\pm0.05^{m}$,
$\overline{M_{H}}=-1.41^{m}\pm0.05^{m}$,
$\overline{M_{Ks}}=-1.52^{m}\pm0.05^{m}$.
The theoretical color dependences of the absolute magnitudes calculated in the same way differ from the empirical ones
obtained for all of the Hipparcos RGC stars located in the regions of the $(B-V)$ -- $M_V$ diagram marked in Fig. 1a.
This difference results from the Hipparcos selection in favor of more luminous stars, among which relatively
blue RGC-Y dominate in our case. This selection also has an effect on the mean absolute magnitudes
but within the above uncertainty of $0.05^{m}$. At the same time, the theoretical and empirical coefficients
of the color dependences differ by up to 30\%, because brighter RGC-Y are simultaneously bluer. Since this
selection should also manifest itself in the Tycho-2 data, as in any magnitude-limited sample of disk
stars, below we use the empirical dependences
\begin{equation}
\label{depend}
M_{B_{T}}=2.89(B_{T}-V_{T})-1.42,~M_{J}=1.25(J-Ks)-1.68.
\end{equation}
The standard deviation of the absolute magnitudes for the RGC stars from the Hipparcos data after applying
a correction for the color dependence is $0.4^m$. This is an estimate of the extent to which we err by replacing
the individual absolute magnitudes with the mean color dependences.

\section*{EXTINCTION AND DISTANCE CALCULATIONS}

Using several photometric magnitudes of a star in the RGC, we can determine the extinction $A_V$, reddening
$E_{(B-V)}$, and unreddened $(B-V)_{0}$ color index for the latter. It is well known that
\begin{equation}
\label{lgr}
5\log(r)-5=B_{T}-M_{B_{T}}-A_{B_{T}}=J-M_{J}-A_{J},
\end{equation}
where r is the distance, $A_{B_{T}}$, $A_J$ are the interstellar
extinctions for the corresponding magnitudes (other magnitudes can also be used). We use the
dependence of extinction $A_{\lambda}$ on the effective wavelength $\lambda$ to reduce all extinctions to the same $A_{V}$.
This dependence is analyzed in detail in many studies, whose results give the difference between the visible
and infrared extinctions with a relative accuracy of better than 5\% used below (see, e.g., Nishiyama et al.
2006; Marshall et al. 2006; and references therein). Having analyzed these data by taking into account
the effective wavelengths $\lambda_{V}=0.553$, $\lambda_{B_{T}}=0.435$, $\lambda_{J}=1.24$ è $\lambda_{Ks}=2.16$ microns,
we adopted
\begin{equation}
\label{btj}
A_{B_{T}}=1.32A_V, A_J=0.28A_V, A_{Ks}=0.11A_V.
\end{equation}
Substituting (1) and (3) into (2) yields an estimate of the interstellar extinction for each star:
\begin{equation}
\label{av}
A_V=(B_{T}-2.89(B_T-V_T)+1.42-J+1.25(J-Ks)-1.68)/1.04.
\end{equation}

For the Hipparcos RGC stars, we found
\begin{equation}
\label{bvbtvt}
(B-V)=0.78(B_T-V_T).
\end{equation}
Taking $E_{(B-V)}=A_V/3$, we calculate the unreddened color index
\begin{equation}
\label{bv0}
(B-V)_0=0.78(B_T-V_T)-A_V/3.
\end{equation}
Negative extinctions were also used. Replacing them by zeroes changes the composition of the sample by less than 1\%.

The photometric distances were calculated from the formula
\begin{equation}
\label{dista}
\log(r)=(5+Ks+1.52-0.11A_V)/5,
\end{equation}
where $r$ is the sought-for distance. We also used negative extinctions, which does not affect the result,
there are no stars with an extinction $A_V<-0.35^{m}$ in the final sample. For the selected stars (see below),
we calculated the Galactic rectangular coordinates $XYZ$, where $X$ increases toward the Galactic center,
$Y$ increases in the direction of Galactic rotation, and $Z$ increases toward the Galactic North Pole.

It should be noted that, in general, the derived extinction and photometric distance are valid only
for the RGC stars due to the replacement of the individual absolute magnitude by its mean color dependence
(1). Therefore, the proper selection of RGC stars discussed below is important.

\section*{THE SELECTION OF STARS}

When the RGC stars are selected from the Tycho-2 Catalogue, it is first necessary to eliminate the
variable and binary stars as well as the stars with inaccurate initial data: we rejected the variables (designated
as G, N in field T47 or U, V, W in Tycho field T48) and binaries (designated as D, R, S, Y, Z in Tycho field T49)
marked in the Tycho Catalogue (ESA 1997, vol. 1, pp. 156--158), including the photocenters of stellar pairs and the
stars with components closer than 25 arcsec (Tycho-2 fields pflag and poflg are not blank, field prox$<250$)
(H\o g et al. 2000), the stars whose $(B_T-V_T)$ had an accuracy lower than $0.15^{m}$, while $(J-Ks)$ and $(J-H)$
had an accuracy lower than $0.05^{m}$, and the stars for which at least one of the proper-motion components had
an accuracy lower than 0.004 arcsec yr$^{-1}$. In addition, imposing reasonable, but fairly wide constraints on the color
indices allowed the binary stars with incorrectly identified components to be rejected.

The RGC stars are usually selected from the 2MASS Catalogue by their positions on the color -- apparent magnitude diagram
(Cabrera-Lavers et al. 2007; and references therein). However, the analysis by Marshall et al. (2006) showed that
distant RGC stars are effectively separated from close MS dwarfs of the same color only in samples of rather faint stars
close to the Galactic plane owing to their distinctly different distances and reddenings: at $V=14^{m}$, an
RGC star with $M_V\approx1$ is located at a heliocentric distance of several kpc and should redden strongly,
as distinct from a MS dwarf with $M_{V}\approx6$ located at a heliocentric distance of $\approx400$ pc.
For the Tycho-2 stars located mostly within the nearest kiloparsec, using the apparent magnitude does not allow RGC
stars to be separated from dwarfs (Marshall et al. 2006).

They can be separated using the reduced proper
motion, which is a substitute for the absolute magnitude
in the absence of systematic stellar motions and
observational selection:
\begin{equation}
\label{mm}
M'_{V}=V+5+5\log(\mu),
\end{equation}
where $\mu=(\mu_{\alpha}^2\cos\delta+\mu_{\delta}^2)^{1/2}$ is the proper motion in arcsecs (Parenago 1954).

The proper motions used to calculate $M'_V$ were corrected for the Galactic rotation and solar motion
to the apex using the photometric distances calculated from (7). For the initial selection of RGC
stars, we adopted typical values for the constants of Galactic rotation
($A=+15$, $B=-10$, $C=0$, $K=0$ km s$^{-1}$ kpc$^{-1}$) and solar motion to the apex
($V=19.5$ km s$^{-1}$, $L=57^{\circ}$, $B=22^{\circ}$) (Parenago 1954).
Subsequently, for the selected stars with radial velocities from the Pulkovo Compilation of Radial Velocities
(PCRV) (Gontcharov 2006), we calculated the velocity components $U$, $V$, $W$ in the Galactic coordinate
system in km s$^{-1}$ and used them to recalculate the constants of Galactic rotation and solar motion
to the apex; subsequently, we corrected $M'_V$ and used them to reselect the stars. After the second iteration,
the calculated parameters of Galactic rotation and solar motion to the apex and the list of selected stars
did not change. The derived constants of Galactic rotation
($A=+15.5\pm2$, $B=-11.3\pm2$, $C=-4.7\pm2$, $K=-3.4\pm2$ km s$^{-1}$ kpc$^{-1}$) and solar motion
to the apex relative to the centroid of the selected stars
($U_{\odot}=+8.4\pm0.5$, $V_{\odot}=+17.6\pm0.3$, $W_{\odot}=+2.7\pm0.4$ km s$^{-1}$, respectively, V = 19.7 km s$^{-1}$,
$L=64.5^{\circ}$, $B=7.9^{\circ}$) agree with those universally accepted for the disk stars.

Figure 1b shows the positions of the same Hipparcos giants and MS stars as those in Fig. 1a but on
the $(B-V)$ -- reduced proper motion $M'_V$ diagram. We see that the RGC is blurred but fairly isolated from
the bright giants, subgiants, and dwarfs. Figures 1c and 1d show similar diagrams for Tycho-2 stars in
a typical sky region before and after dereddening, respectively. We see that there are almost no dwarfs of
the same color as the RGC among the Tycho-2 stars. As in Hipparcos, the RGC in Tycho-2 is identified as
a region of enhanced distribution density of stars. In Fig. 1d, the region within the distribution density isoline
corresponding to the maximum density gradient is shaded. In the same figure, the two ellipses that
describe best this isoline indicate the adopted region all stars of which were selected as presumed RGC
stars and they are considered below. The left and right ellipses roughly reflect the distributions of RGC-Y
and RGC-O, respectively. Comparison of the regions marked in Figs. 1a and 1d shows that the ellipses are
located in the region of theoretically predicted color indices. The maximum of the distribution of RGC
stars is $(B-V)_0=0.92^{m}$, $M'_V=5$. The centers and radii of the ellipses for RGC-Y and RGC-O are
\begin{equation}
\label{oval1}
(B-V)_{0}=0.90^m\pm0.13^m, M'_{V}=4.7\pm2.1,
\end{equation}
\begin{equation}
\label{oval2}
(B-V)_{0}=0.98^m\pm0.20^m, M'_{V}=5.1\pm1.3.
\end{equation}
These quantities do not depend on the Galactic coordinates.

For bright stars ($J<5^{m}$), the 2MASS infrared photometry is inaccurate and many of the bright
(close) RGC stars selected by the described method have erroneous color indices and should be rejected.
This is how a fictitious region of reduced distribution density of stars within about 150 pc of the
Sun emerges. To avoid this, we selected the bright stars directly from Hipparcos when the following
conditions were satisfied: (1) $J<5^{m}$, $H<5^{m}$, $Ks<5^{m}$; (2) the accuracy of the IR magnitudes used is
lower than $0.05^{m}$; (3) $\pi>0.005"$; and (4) $-1.5^{m}<M_{V_{T}}<1.3^{m}$ for $0.75^{m}<(B-V)_{0}<1.02^{m}$ or
$0.5^{m}<M_{V_{T}}<1.5^{m}$ for $1.02^{m}<(B-V)_{0}<1.20^{m}$.
We used the extinction from the model by Arenou et al. (1992). There were 1496 such stars. Below, we use their
trigonometric distances.

We can see how efficient the use of the reduced proper motions for the selection of RGC stars is from
the following Monte Carlo simulations. We specify the distributions of subgiant, RGB, RGC, and dwarf
stars with color indices $0.77^{m}<(B-V)_{0}<1.2^{m}$ in space velocity $V$ and apparent and absolute magnitudes
$V$ and $M_V$. Using the equations
\begin{equation}
\label{rr}
5\log(r)-5=V-M_V,
\end{equation}
\begin{equation}
\label{muvr}
\mu=V/4.74r,
\end{equation}
Eq. (8), and the assumption about an initially uniform selection-biased distribution of stars in space, we
then obtain the distributions of subgiant, RGB, RGC, and dwarf stars from the reduced proper motion $M'_V$.
A detailed discussion of the simulations and the use of the reduced proper motions in various studies deserves
a separate paper. The main simulation results are the following.

G-M dwarfs are almost absent in Tycho-2 and among the selected stars. The main reason is that
this catalogue is magnitude-limited and, accordingly, the volume of space where the dwarfs are selected
is small, no father than 150 pc. Dwarfs probably constitute a majority among the selected stars with
a true distance up to 150 pc, but, according to our simulations, fewer than 500 stars ($<0.5\%$ of the entire
sample; for comparison, 1233 RGC stars within 150 pc were selected directly) should be selected in
this region of space based on photometry and $M'_V$. Since the absolute magnitudes erroneously adopted
for the dwarfs (just as for the RGC stars) are several magnitudes brighter than the true ones, their photometric
distances are larger than the true ones by a factor of several. As a result, these dwarfs should
be a small admixture ($<1\%$) that is rather uniformly distributed in photometric distance and that does not
produce any biases in the results (not the dwarfs, but the RGC stars selected directly dominate even near
the Sun).

RGB stars and subgiants can initially account for up to 20\% of the selected stars. Since the adopted
absolute magnitudes and extinctions do not correspond to the true ones, most of the subgiants and
many of the RGB stars have negative extinctions and photometric distances of more than 500 pc, while
some of the RGB stars have high extinctions at small distances. To reduce the fraction of these stars, we
rejected the stars with $A_{V}>0.6^{m}/(|\sin(b)|+0.18)$, where $b$ is the Galactic latitude. This limit is a reasonable
constraint, since it approximately twice the mean extinction for a given latitude determined from
the model by Arenou et al. (1992) within 1 kpc and is $3.3^{m}$ near the Galactic equator. In addition, because
of the color difference between the RGC stars and subgiants, the latter are largely eliminated by the constraint
$V_{T}-Ks>0.75(B_{T}-J)+0.05$. The remaining subgiants should have photometric distances that
are mostly larger than 500 pc. As a result, according to our simulations, the selected set of presumed RGC
stars within 500 pc should contain less than 10\% of the RGB stars and subgiants and less than 1\% of the
dwarfs.

These conclusions are confirmed by the statistics of the selected stars. For 20 820 selected stars, there
is a spectral classification in the Tycho Spectral Types (TST) Catalogue (Wright et al. 2003), including
the luminosity class. Among them, there are 18 505 (89\%) G5III-M2III giants (the RGC and
part of the RGB, given the classification errors), 281 (1.3\%) GV-MV dwarfs, 1298 (6\%) GIV-KIV
subgiants, 633 (3\%) bright giants and supergiants, and 103 (0.7\%) other stars.

We selected a total of 97 348 presumed RGC stars from about 2.5 million Tycho-2 stars. 14 291 selected
stars are contained in Hipparcos and 45 813 are contained in the TST Catalogue; the spectral classification
of 51 412 selected stars (53\%) has not been known to the present day.

The selected stars have fairly accurate proper motions and photometry: for 85\% of the stars,
$\mu_{\alpha}\cos\delta$ and $\mu_{\delta}$ are known to within 0.003 arcsec yr$^{-1}$, $B_T$ and
$V_T$ are known to within $0.15^{m}$ and $0.1^{m}$, respectively,
and the infrared magnitudes are known to within $0.05^{m}$.

The selection ellipses specified by conditions (9) and (10) contain approximately equal numbers of
stars. However, the fraction of RGC-Y and RGC-O is difficult to estimate accurately, since the zone
of intersection between the ellipses is large and, in addition, the admixture of foreign stars is distributed
nonuniformly. Only the fact that RGC-Y account for a significant fraction (possibly up to half) of the RGC
stars is beyond question. Below, we consider various characteristics of the selected stars by dividing the
sample into two equal parts, depending on whether a star is closer to the center of one or the other ellipse
on the $(B-V)_0$ -- $M'_V$ diagram. Below, these groups of stars are arbitrarily called RGC-Y and RGC-O.

\section*{THE DISTRIBUTION OF RGC STARS IN SPACE}

Figure 2 shows the distribution of the selected stars on the celestial sphere in Galactic coordinates.
The distributions of RGC-Y and RGC-O are identical. Regions of reduced density, ``voids'', are clearly
seen not far from the Galactic equator or, more precisely, along the great circle inclined to the equator by
about $15^{\circ}$ and coincident with the equatorial plane of the Gould Belt. Obviously, these voids result
from relatively strong absorption by the matter located within several hundred parsecs of the Sun in layers parallel
to the equatorial plane of the Gould Belt.

Figure 3a compares the photometric distances with the distances calculated from the Hipparcos
parallaxes for 13 079 stars with $\pi>0.001$ arcsec (RGC-Y and RGC-O are shown together). We see agreement
between the distances within 350 pc (apart from the stars selected directly), where the Hipparcos
parallaxes are fairly accurate. The ratio of the standard deviation of the difference between the photometric
and astrometric distances to the astrometric distance, $\sigma(r_{ph}-r_{HIP})/r_{HIP}$, for the Hipparcos stars with
350 pc is 0.35.

As we see from Fig. 3a, the photometric distances for some of the stars are appreciably smaller than
the distances determined from the Hipparcos parallaxes. For these stars, the extinction calculated here
is considerably lower than that corresponding to the Hipparcos parallax. This can be seen from Fig. 3b,
which compares the derived extinction $A_V$ with that calculated by Arenou et al. (1992) using the Hipparcos
parallaxes (RGC-Y and RGC-O are shown together). Good agreement for most of the stars suggests
that the proposed method of correction for the extinction is applicable.

Since the reduced proper motions were used here only for the selection of stars, but not for the
calculation of distances, in contrast to our previous study (Gontcharov 2008), the derived distances will
not have the corresponding systematic errors. In addition, the systematic errors that arise when the
trigonometric parallaxes are used to calculate the distances do not arise here either. The accuracy of
the derived distances is determined by the accuracy of the initial photometric quantities, the spread in
absolute magnitude for the RGC stars, the fraction of stars with a different absolute magnitude, the
error in the adopted absorption coefficient, and other factors. The estimates of these uncertainties made
previously allow us to estimate the relative accuracy of the derived photometric distances, about 25\%
(corresponding to a photometric error of 0.5$^{m}$). Thus, for distances larger than 350 pc, the photometric
distances should be more accurate than the distances determined from the Hipparcos parallaxes.

In Fig. 4, the extinction $A_V$ is plotted against the distance for (a) RGC-Y and (b) RGC-O. We see the
mean extinction to increase up to a distance of about 550 and 450 pc, respectively (the right white dotted
line). Further out, the mean extinction decreases. This results from selection, because the catalog is
magnitude-limited (Gontcharov 2008). Far from the Galactic plane, the RGC-Y and RGC-O subsamples
are selection-unbiased up to these distances. Near the Galactic plane, the corresponding distance is determined
from the maximum extinctions, which occur at a distance of about 350 pc marked by the left white
dotted line for both subsamples. As expected, RGC-Y are seen at larger heliocentric distances because of
their higher mean luminosities. There are more stars with negative extinctions among RGC-O, which is
due probably to the larger admixture of RGB stars and subgiants.

Jointly using the 2MASS and Tycho-2 Catalogues allows a sample of RGC stars to be considered
in a larger region of space than that using Hipparcos. This can be seen from Fig. 5, which shows the spatial
distribution of stars in projection onto the $XY$, $XZ$, and $YZ$ planes: (a) calculated from the Hipparcos
parallaxes for 13 889 selected presumed RGC stars with positive Hipparcos parallaxes; (b) calculated
from the photometric distances all 97 348 selected RGC stars; (c) the same indicated by a contour map
for half of the selected conditional RGC-Y; (d) the same for conditional RGC-O.

We see that the Hipparcos RGC stars are distributed uniformly in space within about 200 pc. It
is virtually impossible to analyze the distribution of RGC stars outside this volume on the basis of Hipparcos
data. In contrast to Hipparcos, our sample of presumed RGC stars reveals a decrease in the
distribution density of stars with distance from the Galactic plane (the plots in the $XZ$ and $YZ$ planes),
with this decrease being slightly different for RGC-Y and RGC-O. On the same plots, we see the role of extinction
in ``withdrawing'' the stars near the Galactic plane; in the first and second Galactic quadrants, this
role is appreciably greater than that in the third and fourth ones, which manifests itself in the mean $Y$ of
the stars under consideration: $\overline{Y}=-25\pm1$ pc. The positions of the ``holes'' agree with the orientation of
the Gould Belt.

Our sample of stars is appreciably more extended along the $Y$ axis than along the $X$ axis, which is
completely invisible in the Hipparcos sample. The Local Spiral Arm of the Galaxy stretching along the
$Y$ axis probably manifests itself in this way. In the Galactic center--anticenter direction, the Sun is located
almost at the center of the spatial distribution of the sample of stars under consideration: $\overline{X}=13\pm1$ pc.
Given a noticeable decrease in density in these directions at a distance of more than 500 pc, which is
probably caused not by selection, but by the location outside the Local Arm, it may be concluded that the
Sun lies approximately at the center of the Local Arm section.

The distribution density of 39 927 selected presumed RGC stars in the cylindrical region of space
$(X^{2}+Y^{2})^{1/2}<400$ pc does not depend on $X$ and $Y$, but depends only on $Z$. Thus, we are probably dealing
with a selection-unbiased sample of RGC stars in the vertical cylindrical section of the Galactic disk in the
solar neighborhood. The mean value of Z for the stars in the cylinder of space under consideration suggests
that the Sun rises above the Galactic plane by $13\pm1$ pc. This also manifests itself in the asymmetric distribution
of stars in the cylinder under consideration along the $Z$ axis, which is shown in Fig. 6 by the
columns for each 100 pc for (a) conditional RGC-Y and (b) RGC-O: at all of the distances, except
$|Z|\approx700$ and $|Z|>900$ pc, the number of stars in each column with positive $Z$ is smaller than that in
the column with $-Z$.

Theoretically, the distribution density of the stars can decrease with distance Z as the square of the
hyperbolic secant $sech^{2}(2Z/Z_{0})$ (Girardi et al. 2005) or according to the barometric law $D_0\cdot~e^{-|Z|/Z_0}$,
where $D_0$ is the density in the equatorial plane, which does not necessarily coincide with the Galactic
plane, and $Z_0$ is the distance from this plane at which the density decreases by a factor of $e$ or
the half-thickness of a homogeneous layer of stars (Parenago 1954, p. 264). The solid curves in Figs. 6a
and 6b indicate the functions $4000e^{-|Z+13^{\circ}|/220}$ and $4900e^{-|Z+13^{\circ}|/230}$ and the dotted curves
indicate $13800/(e^{(Z+13^{\circ})/280}+e^{-(Z+13^{\circ})/280})$ and
$16800/(e^{(Z+13^{\circ})/280}+e^{-(Z+13^{\circ})/280})$, whose coefficients were found by the least-squares method.
The distributions of RGC-Y and RGC-O are better described by the barometric law and the square of
the hyperbolic secant, respectively. The values of $Z_0$ found are typical of the Galactic thin-disk stars.

\section*{THE KINEMATICS OF RGC STARS}

For 4163 selected stars with radial velocities from the PCRV and RAVE catalogs (Steinmetz et al. 2006), we calculated
the velocity components $U$, $V$, $W$ in the Galactic coordinate system in km s$^{-1}$.
These stars include the 1990 stars considered by Famaey et al. (2005), but, in contrast to them,
they are distributed fairly uniformly over the sky. These data allow some general conclusions about the
kinematics of the selected presumed RGC stars to be reached (a detailed analysis will be presented in a
separate publication).

Above, we mentioned the parameters of Galactic rotation and solar motion to the apex calculated from
these data and consistent with those universally accepted for disk stars.

Figure 7 shows the distribution of the 4163 stars under consideration in velocity components $U$ and $V$:
(a) the 900 stars closest to the center of the RGC-Y ellipse and farthest from the center of the RGC-O
ellipse on the $(B-V)_{0}$ -- $M'_{V}$ diagram, i.e., presumably the youngest and most massive stars of the sample
and (b) the remaining stars. Comparison with the results of Famaey et al. (2005) suggests that subsample
(a) actually contains stars younger than 1 Gyr, while subsample (b) contains typical disk stars of
various ages. Thus, it is confirmed that the RGC contains a significant fraction (probably $>20\%$) of
relatively young massive stars.

For both RGC-Y and RGC-O, the dispersion of the velocity component W increases with distance
from the Galactic plane, on average, from 18 km s$^{-1}$ in the plane to 30 km s$^{-1}$ at $|Z|=500$ pc. The deviations
from a monotonic change in dispersion observed at some $Z$ require a separate study.

We found contraction of the set of stars under considerations along the $Z$ axis, i.e., a velocity gradient
$W/Z=-16\pm4$ è $W/Z=-12\pm4$ km s$^{-1}$ kpc$^{-1}$ for RGC-Y and RGC-O, respectively.

\section*{CONCLUSIONS}

This study has shown that the 2MASS and Tycho-2 broad band multicolor photometry and proper
motions are sufficient to calculate the individual interstellar extinctions and photometric distances for
RGC stars. Using the method under consideration, we selected the stars in a region of space that exceeded
a similar region for the Hipparcos Catalogue by a factor of several. Within about 350 pc of the
Sun, the sample will be selection-unbiased. This allowed us to detect a decrease in the distribution
density of stars both with height above the Galactic plane and at the presumed edges of the Local Spiral
Arm and to determine the rise of the Sun above the Galactic plane with a high accuracy. In the region
of space under consideration, the RGC was shown to be a mixture of three groups of stars, including a
significant fraction of stars younger than 1 Gyr with masses higher than 2 $M_{\odot}$.

The method under consideration is important for investigating Galactic structures and for modern
large astronomical projects, whose key objective is an automatic classification of millions of stars and extended
objects based on their multicolor photometry.

\section*{ACKNOWLEDGMENTS}

I wish to thank the referee for helpful remarks. In this study, I used the 2MASS (Two Micron All
Sky Survey) Catalogue, which is a joint project of the Massachusetts University and the IR Data
Reduction and Analysis Center of the California Institute of Technology financed by NASA and
the National Science Foundation. I also used resources from the Strasbourg Data Center (France)
(http://cdsweb.u-strasbg.fr/). This study was supported by the Russian Foundation for Basic Research
(http://www.rfbr.ru) (project nos. 05-02-17047 and 08-02-00400).

\newpage

\begin{figure}
\includegraphics{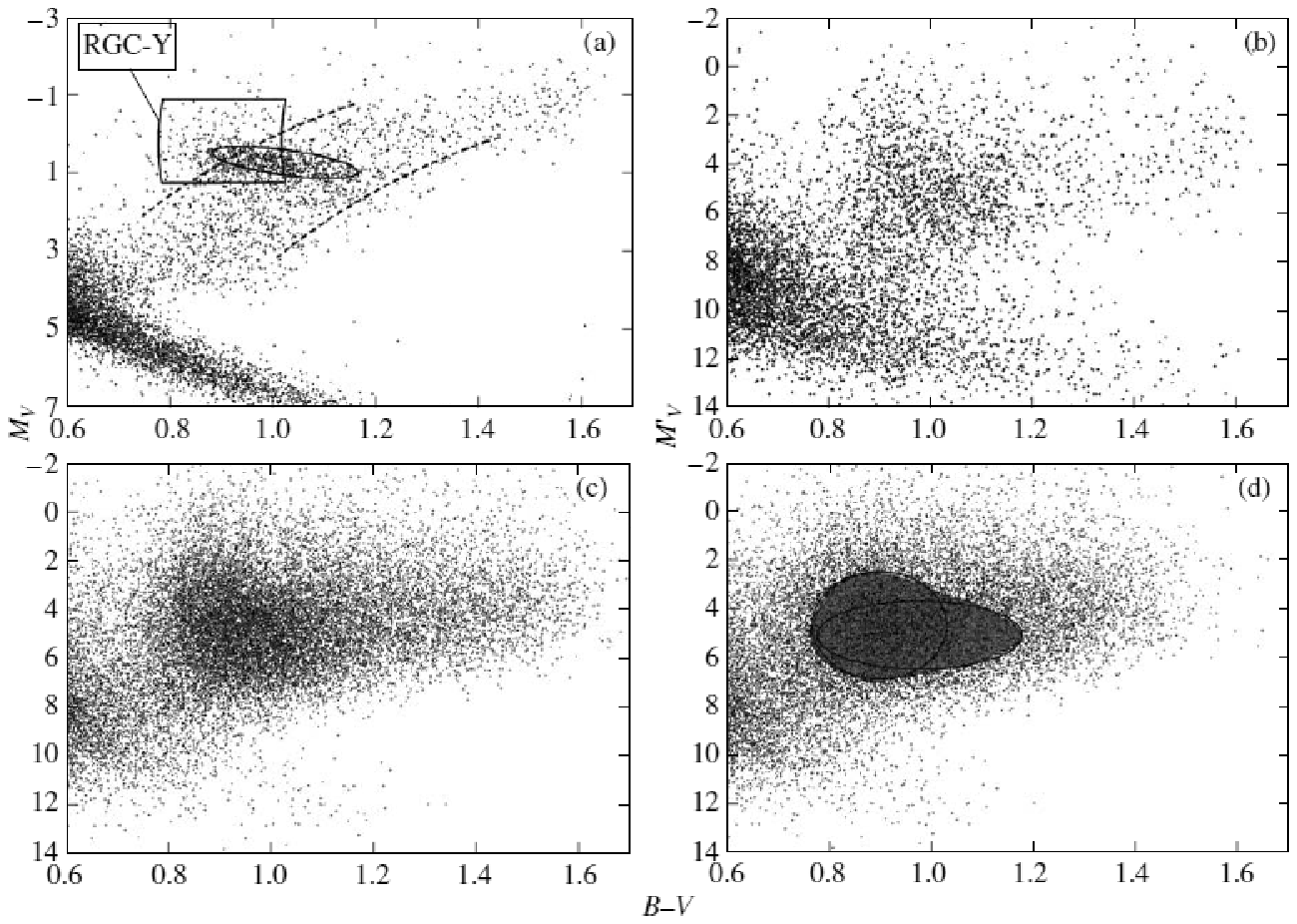}
\caption{Giants and part of the MS (a) on the $(B-V)$ -- $M_V$ diagram for Hipparcos stars with the best data,
(b) on the $(B-V)$ -- reduced proper motion $M'_V$ diagram for the same Hipparcos stars, (c) on the $(B-V)$ -- $M'_{V}$
diagram for Tycho-2 stars in a typical sky region before dereddening, and (d) the same after dereddening.
The shading marks the region of enhanced distribution density of stars; the two ellipses mark the RGC star selection region.
}
\label{hr}
\end{figure}


\begin{figure}
\includegraphics{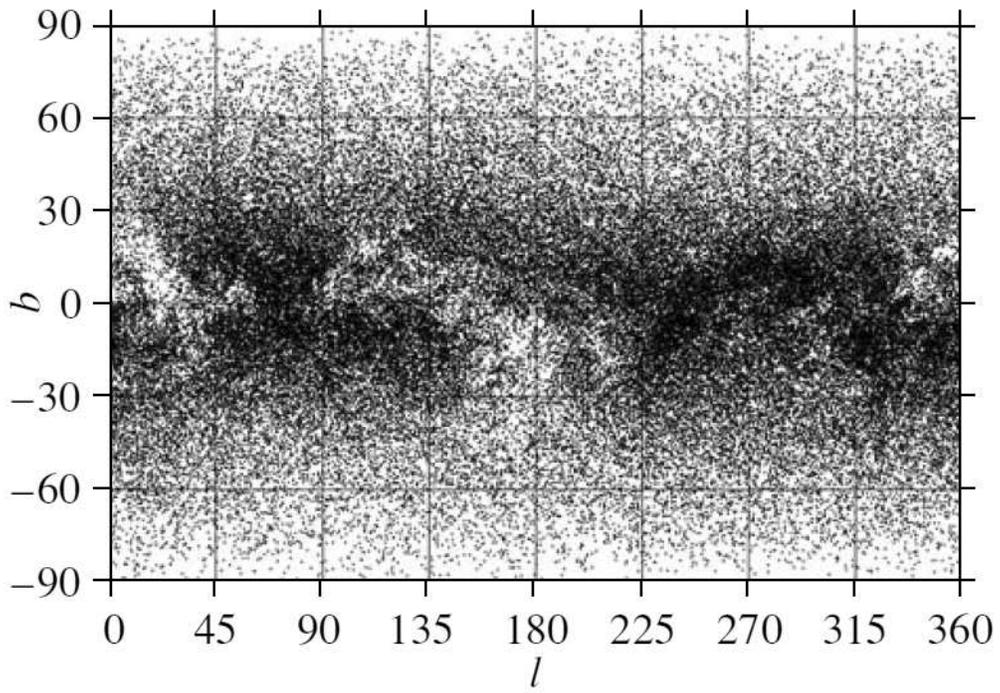}
\caption{Distribution of 97 348 selected presumed RGC stars on the celestial sphere in Galactic coordinates.
}
\label{lb}
\end{figure}


\begin{figure}
\includegraphics{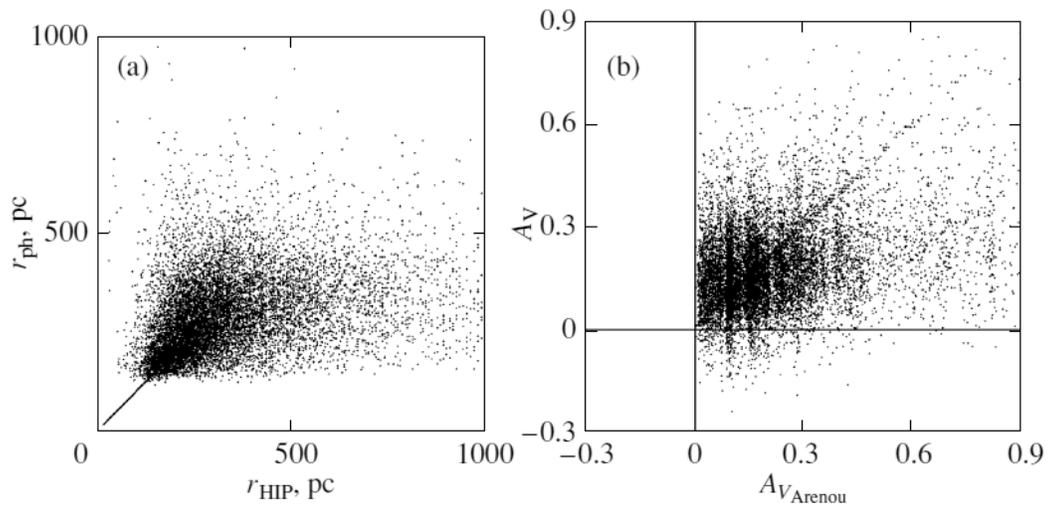}
\caption{Comparison of the (a) photometric distances with the distances calculated from the Hipparcos parallaxes and
(b) the derived extinction with that calculated by Arenou et al. (1992) from the Hipparcos parallaxes for 13 079
selected presumed RGC stars with $\pi>0.001$ arcsec.
}
\label{rraa}
\end{figure}


\begin{figure}
\includegraphics{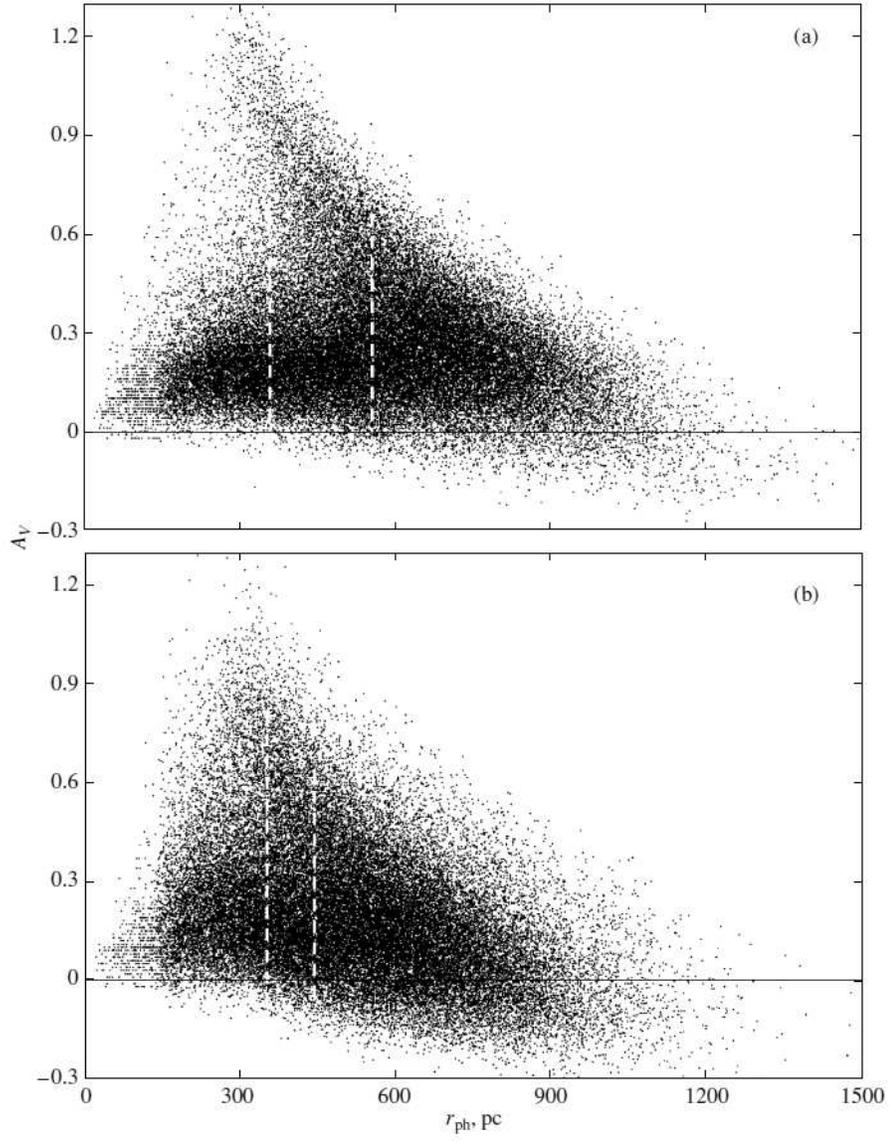}
\caption{Interstellar extinction $A_V$ versus photometric distance for (a) 48 674 selected conditional RGC-Y and
(b) conditional RGC-O.
}
\label{avr}
\end{figure}


\begin{figure}
\includegraphics{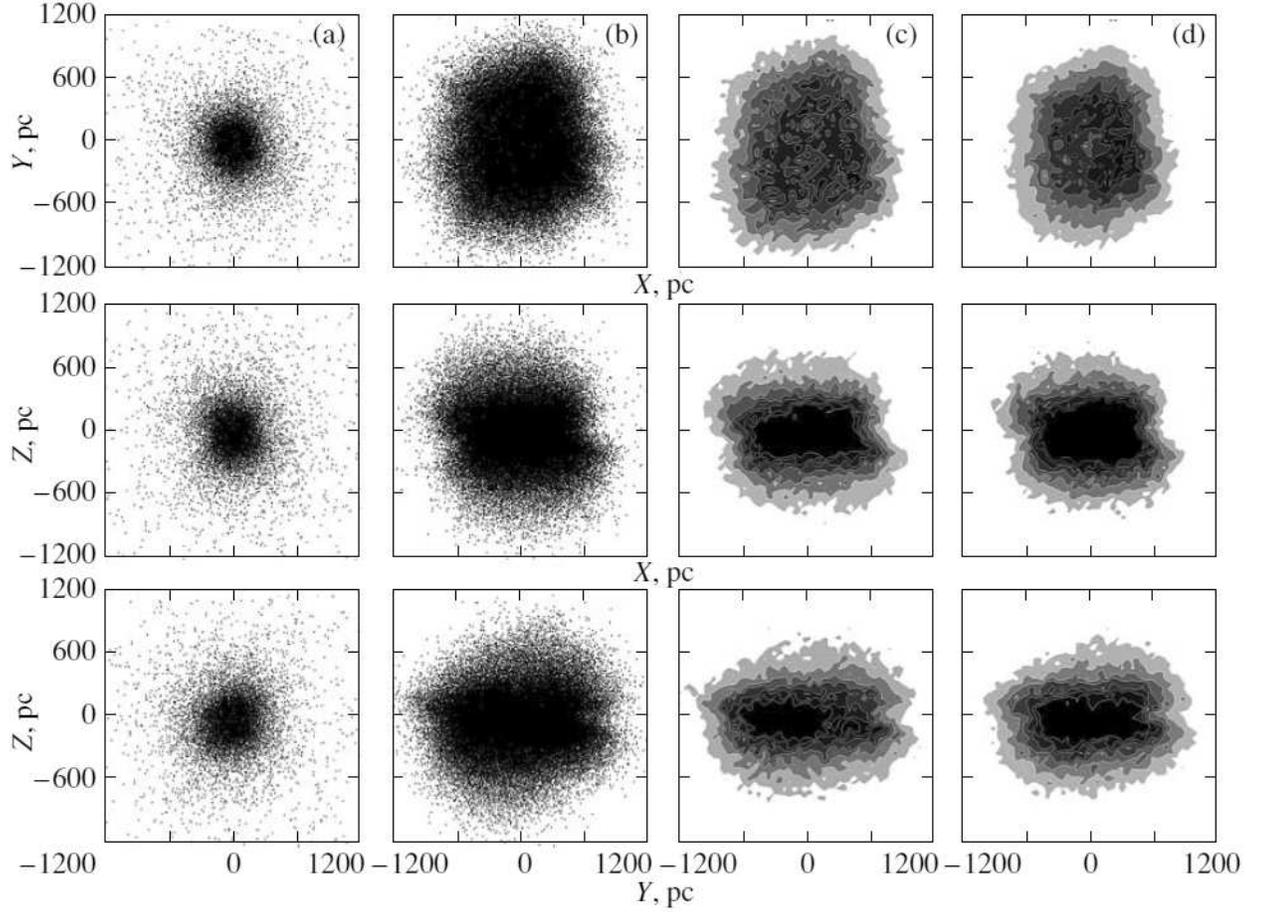}
\caption{Spatial distributions in projection onto the $XY$, $XZ$, and $YZ$ planes calculated
(a) from the Hipparcos parallaxes for 13 889 selected presumed RGC stars with positive Hipparcos parallaxes,
(b) from the photometric distances for all 97 348 selected presumed RGC stars,
(c) the same indicated by a smoothed contour map for 48 674 conditional RGC-Y, and
(d) the same indicated by a smoothed contour map for 48 674 conditional RGC-O.
}
\label{xyz}
\end{figure}


\begin{figure}
\includegraphics{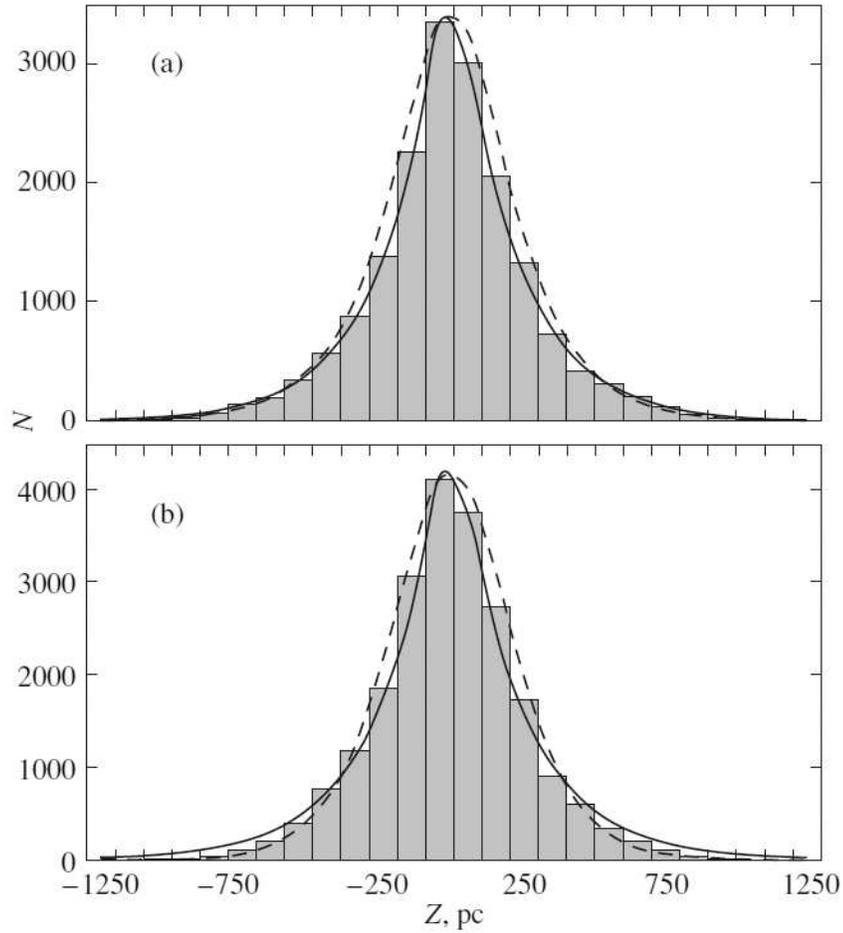}
\caption{Distributions of (a) 17 623 selected conditional RGC-Y and (b) 22 304 RGC-O in the vertical cylinder along
the $Z$ axis within 400 pc of the Sun; the solid and dotted curves indicate the variations in distribution density
according to the barometric law and as the square of the hyperbolic secant, respectively.
The asymmetry results from the rise of the Sun above the Galactic plane.
}
\label{zn}
\end{figure}


\begin{figure}
\includegraphics{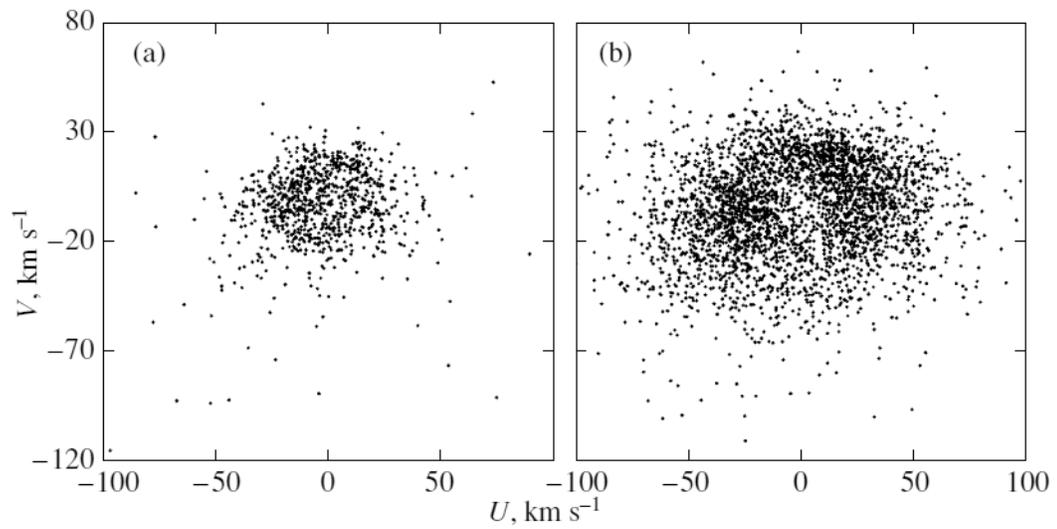}
\caption{Distribution of the 4163 stars under consideration with radial velocities derived from the velocity components
$U$ and $V$:
(a) the presumed 900 youngest and most massive stars and (b) the remaining stars.
}
\label{uv}
\end{figure}

\end{document}